\documentstyle[aps,epsf]{revtex}

\newcommand{\be}{\begin{equation}}
\newcommand{\ee}{\end{equation}}
\newcommand{\bea}{\begin{eqnarray}}
\newcommand{\eea}{\end{eqnarray}}

\newcommand{\Eq}[1]{Eq.~(\ref{#1})}

\begin{document}

\draft

\title{Life and Death near a Windy Oasis}
\author{Karin A. Dahmen, David R. Nelson}
\address{Lyman Laboratory of Physics,
Harvard University, Cambrigde, MA, 02138}
\author{Nadav M. Shnerb}
\address{Racah Institute of Physics, Hebrew University, Jerusalem 91904,
Israel}

\maketitle

\begin{abstract}
We propose a simple experiment to study delocalization
and extinction in inhomogeneous biological 
systems.
The nonlinear steady state for, say, a bacteria colony living on 
and near 
a patch of nutrient or favorable illumination (``oasis'') 
in the presence of a drift term 
(``wind'') is computed. The bacteria, described by a simple
generalization of the Fisher equation, diffuse,
divide $A \rightarrow A+A$, die $A \rightarrow 0$, 
and annihilate $A+A \rightarrow 0$.
At high wind velocities all bacteria are blown into an unfavorable
region (``desert''), and the colony dies out.
At low velocity a steady state concentration
survives near the oasis. In between these two regimes
there is a critical velocity at which bacteria first survive.
If the ``desert'' supports a small nonzero population,
this extinction transition is replaced by a delocalization 
transition with increasing velocity.
Predictions for the behavior as a function of wind velocity
are made for one and two dimensions.
\end{abstract}
%
%
%
\section{Introduction and Results}
Bacterial growth in a petri dish, the basic experiment of microbiology,
is a familiar but interesting phenomenon. Depending on the nutrient 
and agar concentration, a variety of intriguing growth patterns 
have been observed\cite{Wakita,Rauprich,Ben-Jacob,Budrene}.
Some regimes can be modeled by diffusion limited aggregation,
others by Eden models, and still others exhibit ring structures or a
two-dimensional modulation in the bacterial density. At high 
nutrient concentration 
and low agar density, there is a large regime of simple growth 
of a circular patch (after point innoculation), described 
by a Fisher equation\cite{murray},
and studied experimentally in Ref.\cite{Wakita}.

Of course, most bacteria do not live in petri dishes, but rather in 
inhomogeneous environments characterized by, e.g., spatially varying
growth rates and/or diffusion constants.
Often, as in the soil after a rain storm
(or in a sewage treatment plant), bacterial diffusion
and growth are accompanied by convective drift 
in an aqueous medium through the disorder. By creating artificially
modulated growth environments in petri dishes, one can begin to 
study how bacteria (and other species populations) grow in circumstances
more typical of the real world. More generally, the challenges posed
by combining inhomogeneous biological processes with various types of 
fluid flows\cite{Robinson} seem likely to attract considerable interest 
in the future. The easiest problem to study in the context of bacteria 
is to determine how fixed spatial inhomogeneities and convective flow
affect the simple regime of Fisher equation growth
mentioned above.

\begin{figure}
\epsfxsize=4truein \vbox{\hskip 1.2truein
\epsffile{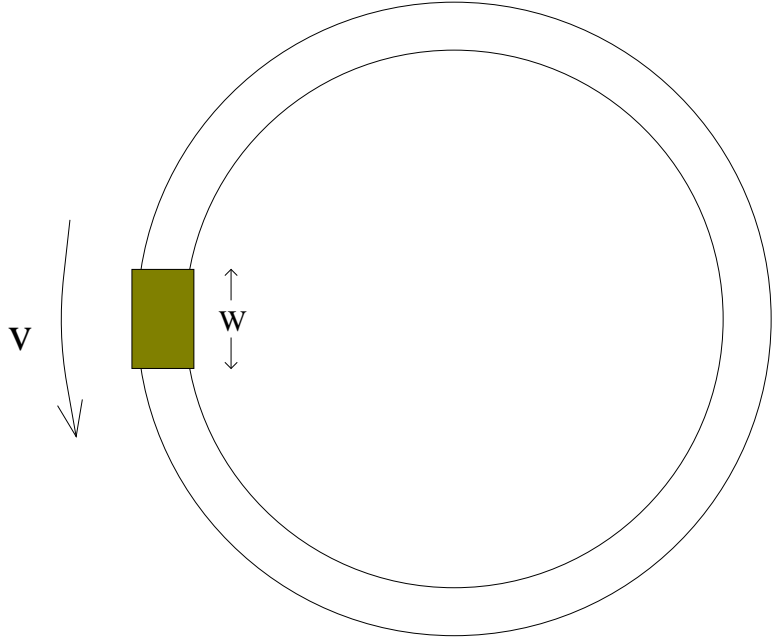}}
\vskip 0.2truein
\caption{Experimental setup:
a solution with photosynthetic bacteria
in a circular pipe or a thin annular track in a petri dish, 
is illuminated only in a small area, while
the rest of the sample is either kept dark or illuminated
with reduced intensity. The light source is moved
slowly around the sample to model convective
flow. The bacteria are assumed to divide in 
the illuminated area (``oasis'') at a certain growth rate $a > 0$, 
and die (or grow modestly) in the remaining 
area (``desert'') with growth rate $ - \epsilon a $.
\label{setup}}
\end{figure}

\begin{figure}
\epsfxsize=5truein \vbox{\vskip 0.15truein\hskip 1truein
\epsffile{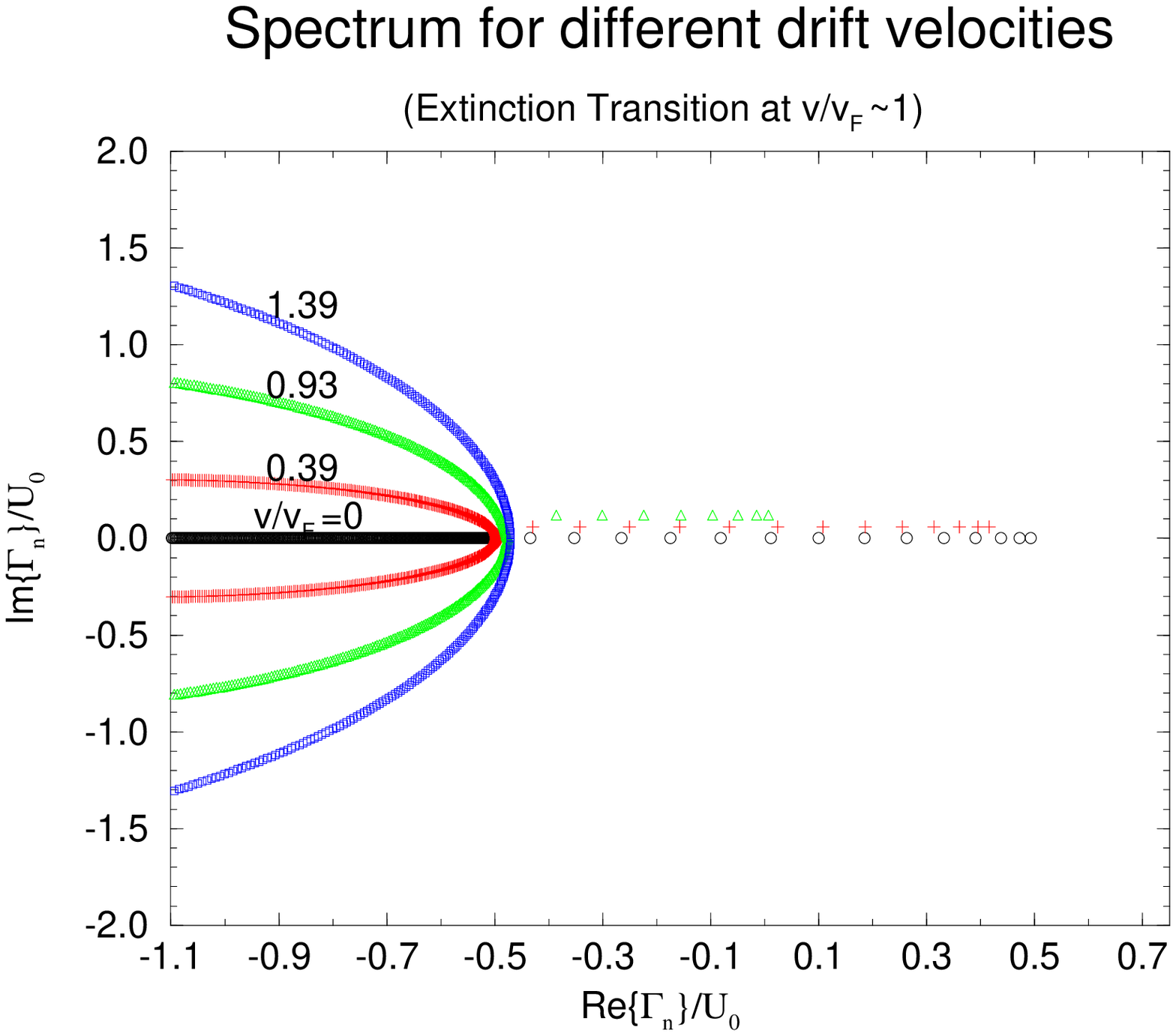}}
\medskip
\caption{Complex nonhermitian eigenvalue spectra 
(normalized by the difference of the growth rates inside
and outside the oasis $U_0=a+\epsilon a = 1$) 
at velocities above and below the extinction transition.
The spectra are extracted from numerical simulations of the 
lattice model described in Appendix {\protect{\ref{app:lattice}}}, 
for a system of 1000 sites, oasis width $W=20$ sites, $D=0.3$, with
growth rate $U=-0.5$ in the desert and $U=+0.5$ in the oasis, 
so that the average growth rate is 
is $-0.48$ (which is equal to the mobility edge
$\Gamma^*$ up to finite size effects).
The chosen velocity parameters 
(in units of the Fisher wave velocity in the oasis 
$v_F= 2\sqrt{aD}$ {\protect\cite{murray}}) are $v/v_F\sim 0$ (circles), 
$v/v_F\sim 0.39$ (plus), $v/v_F\sim 0.93$ (triangles), 
and $v/v_F\sim 1.39$ (squares).
The point spectra are slightly offset in the $y$-direction so as to
be able to distinguish the eigenvalues of the localized states
for different velocities. As described in the introduction,
the mobility edge remains roughly fixed, and the parabola
of the delocalized eigenvalues opens up as $v/v_F$ 
is increased. The real, localized eigenvalues move to the left for
higher velocities. At $v/v_F \geq 1.39$ all states are delocalized.
(This figure actually only shows the part of the spectrum 
which corresponds to the continuum problem. The lattice 
calculation also yields a left part of the spectrum 
-- not shown here -- which is an artifact of the discrete lattice.)
\label{Gamma-v-series}}
\end{figure}

A delocalization transition in inhomogeneous biological systems
has recently been proposed, 
focusing on a single species continuous growth model, in which 
the population disperses via diffusion and convection\cite{nadav}:
the Fisher equation\cite{murray} for the population number density 
$c({\mathbf x},t)$, generalized to account for convection and 
an inhomogeneous growth rate, reads \cite{nadav,continuous}
\bea
\label{equation-of-motion}
\nonumber
\partial c({\mathbf x},t)/\partial t &=&
D \nabla^2 c({\mathbf x}, t) - {\mathbf v} \cdot \nabla c({\mathbf x},t) \\
& & + U({\mathbf x}) c({\mathbf x}, t)
- b c^2({\mathbf x}, t) \, ,
\eea
where $D$ is the diffusion constant of the system,
${\mathbf v}$ is the spatially homogeneous convection (``wind'') 
velocity, and
$b$ is a phenomenological parameter responsible for the limiting of the 
concentration $c({\mathbf x}, t)$ to some maximum saturation value
(by competition processes of the kind $A+A \rightarrow 0$\cite{continuous}).
The growth rate $U({\mathbf x})$ is a random function
which describes a 
spatially random nutrient concentration, or, for photosynthetic 
bacteria, an inhomogeneous illumination pattern \cite{nadav}.
If $U({\mathbf x})$ is constant over the entire sample, then the convection
term $- {\bf v} \cdot \nabla c({\bf x}, t)$ has no effect on the growth
of the bacteria. Only the introduction of a spatial dependence
for the growth rate $U({\mathbf x})$ makes the convection term interesting.
In the following we consider the simple case of a
``square-well potential'' shape for $U({\mathbf x})$,
imposing a positive
growth rate $a$ on an illuminated patch (``oasis''), 
and a negative growth rate $ - \epsilon a $
outside (``desert'') \cite{Ben-Avrahams}: 
\be
\label{V-def}
%
U({\mathbf x}) = 
\cases{
a\, ,& for $|{\mathbf x}|<{W\over2}\, ,$\cr
\ & \cr
-\epsilon a\, ,& for $|{\mathbf x}|\geq{W\over2}\, ,$\cr
}
\ee
where $W$ is the diameter of the oasis.
Experimentally 
this could be realized using  a very simple
setup, which both illustrates the basic ideas of localization
and delocalization, and leads to interesting further questions.
A one dimensional example is shown in
figure~\ref{setup}, where a solution with photosynthetic bacteria
in a thin circular pipe, or annular petri dish,
is illuminated by a fixed uniform light source through a mask,
leading to a ``square well'' intensity distribution.
The mask is moved at a small, 
constant velocity around the sample to simulate convective
flow. (Moving the mask is equivalent to introducing convective flow 
in the system, up to a change of reference frame\cite{ref-frame}.)
The bacteria are assumed to divide in the brightly 
illuminated area (``oasis'') at a certain rate, but division 
ceases or proceeds at a greatly reduced rate in the darker 
region (``desert'') outside. 
As a result, the growth rate in this continuum population 
dynamics model is positive in the oasis and  
small (positive or negative) in the surrounding desert region. 
Using this simple nonlinear growth model, we discuss predictions for 
the total number of bacteria expected to survive in the steady state,
the shape of their distribution in space and other quantities,
as a function of the ``convection velocity'' of the light source.
\
It is interesting to consider the class of biological situations 
discussed in this paper in the context of the 
"critical size problem"  in population  dynamics \cite{okubo}. 
In the critical size problem one asks for the 
minimal size of habitat for the survival of a population
undergoing logistic growth and diffusion, where the region is surrounded
by a totally hostile environment, i.e., 
no drift and an  infinite death rate outside the oasis. We show 
here that the linearized version of the critical size 
problem is closely related to a well known 
problem in quantum mechanics, and
present a generalization of this problem to include other types of 
surrounding environments, as well as the effect of drift. 
The linearized version of 
equation~(\ref{equation-of-motion}) around $c({\mathbf x},t)=0$
reads
\be
\label{linearized-equation}
\partial c({\mathbf x},t)/\partial t = {\cal L} c({\mathbf x},t) \, ,
\ee
with the linearized growth operator 
\be
\label{Liouville}
{\cal L}= D \nabla^2 - {\mathbf v} \cdot \nabla + U({\mathbf x}) \, .
\ee
(We discuss later the validity of this linear approximation and 
compare the results with lattice simulations of the full nonlinear
problem.)
For nonzero convection velocity $v$, ${\cal L}$ is non-Hermitian,
but it can still be diagonalized by a complete set of right and left
eigenvectors, $\{ \phi^R_n({\mathbf x}) \}$ and 
$\{ \phi^L_n({\mathbf x}) \} $,
with eigenvalues $\Gamma_n$ \cite{Hatano,nadav},
and orthogonality condition
\be
\int d^dx \phi^L_m({\mathbf x}) \phi^R_n({\mathbf x}) = \delta_{m,n} \, ,
\ee
($d$ is the dimension of the substrate, we focus here on
$d=1$ or $d=2$).
The time evolution of $c({\mathbf x}, t)$ is then given by
\be
\label{time-evolution}
c({\mathbf x}, t) = \sum_n c_n \phi^R_n({\mathbf x}) \exp(\Gamma_n t) \, ,
\ee
where the initial conditions and left eigenfunctions
determine the coefficients
$\{c_n\}$,
\be
c_n=\int d^dx \phi^L_n({\mathbf x}) c({\mathbf x}, t=0) \, .
\ee
Figure \ref{Gamma-v-series} shows the complex
eigenvalue spectrum with the potential (\ref{V-def})
for four different values of the convection velocity $v$, for a
one dimensional lattice approximation to (\ref{Liouville}) 
(see Appendix \ref{app:finite-syst}) with periodic boundary 
conditions. The derivation of these
results is discussed in section \ref{linear} below.
At zero velocity $\cal{L}$ is Hermitian and all eigenvalues $\Gamma_n$
are real. There are bound states (discrete spectrum) and 
extended or delocalized states (continuous spectrum).
At finite velocities, all except one of the 
delocalized states acquire a complex
eigenvalue. States with positive real part of the eigenvalue
($Re{\Gamma_n}>0$) grow exponentially with time, states with negative 
real part ($Re{\Gamma_n}<0$) decrease exponentially with time 
(see \Eq{time-evolution}).
In a large one dimensional system 
the ``mobility edge''\cite{mobility-edge}, which 
we define to be the eigenvalue of the fastest growing delocalized state,
(i.e. the rightmost eigenvalue in the 
complex parabolas of figure \ref{Gamma-v-series}),
is located at the overall average growth rate 
\be
\Gamma^* = \langle U \rangle \equiv \int_0^L dx U(x)/L \simeq -\epsilon a \, ,
\ee
(up to corrections of order $O(1/L)$ where the system size $L$ 
is the mean circumference of the annulus in figure \ref{setup}).
In figure \ref{Gamma-v-series},
the eigenvalues of the localized states compose the discrete, real
spectrum to the right of the mobility edge.
With increasing velocity these localized eigenvalues move
to the left by an amount proportional to $v^2/4D$, and 
successively enter the continuous delocalized spectrum
through the mobility
edge, which remains fixed. The parabola broadens in the 
vertical direction -- the imaginary parts of the eigenvalues 
of the delocalized states grow by an amount proportional to $v$.
A given localized right eigenfunction $\phi_n^R$
undergoes a ``delocalization transition''
when the velocity reaches a corresponding critical delocalization
velocity $v=v^*_n$, at which its eigenvalue $\Gamma_n$ has been
shifted so far to the left that it just touches $\Gamma^*$.
At higher velocities it joins the parabola of eigenvalues 
describing a continuum of delocalized states.
The ease with which such a delocalization transition can be observed
experimentally depends on whether there are 
growing delocalized eigenstates in the system, i.e. whether the
mobility edge has a positive real value or not.
In a large ``deadly'' desert 
($\langle U \rangle \simeq -\epsilon a < 0 $)
all delocalized states die out,
because the mobility edge lies to the left of the origin,
as in figure \ref{Gamma-v-series}.
The growth rate of each localized eigenstate $\phi^R_n$ then
becomes negative at a corresponding ``extinction'' velocity $v_{nc}$
which is smaller than the corresponding delocalization velocity
$v^*_n$. Thus, as convection is increased, the population
dies out before it can delocalize. 
Later in this paper, we make specific predictions
for the behavior of populations near the extinction transition,
which occurs for $v=v_{0c} > v_{nc}$ for all $n>0$,
when the eigenvalue of the localized
``ground state'' (fastest growing eigenfunction of $\cal{L}$)
passes through the origin.

If the average growth rate $\langle U \rangle$ is positive
({\it i.e} for a small enough desert or a small positive growth rate in an
infinite desert),
the mobility edge lies to the right of the origin and the 
delocalization transition 
can indeed be observed at $v=v_0^*$ where the ``ground state'' becomes
delocalized.
One expects to see {\it universal} behavior near
this delocalization transition, since there is a diverging correlation
length in the system, which renders microscopic details irrelevant
for certain quantities. We report predictions 
(see also \cite{long-paper}), for quantities such as
the dependence of the localization length on the drift velocity 
as it approaches the delocalization velocity, and the shape
of the concentration profile near the transition.

\begin{figure}
\epsfxsize=5truein \vbox{\vskip 0.15truein\hskip 1truein
\epsffile{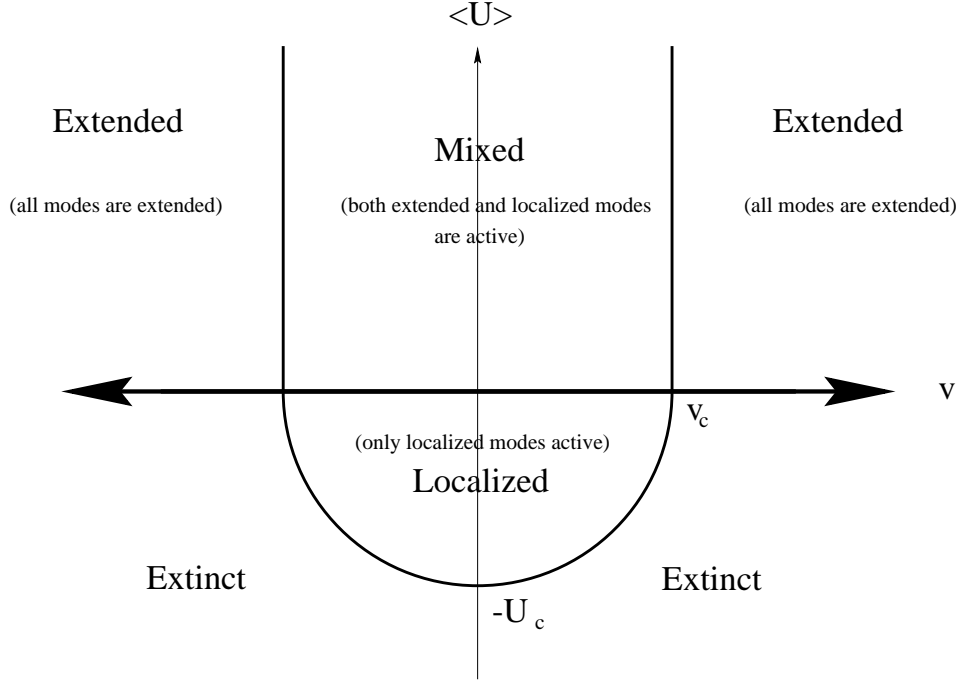}}
\medskip
\caption{Schematic phase diagram in one dimension
for infinite system size, as a function of average growth
rate $\langle U \rangle$ and convection velocity $v$.
For a deep well ($U_0 \equiv a + \epsilon a \gg D/W^2$)
the extinction transition out of the localized phase occurs
when $v=0$ for $\langle U \rangle = - U_c$, where $U_c \simeq U_0$.
The diagram shows that if the growth rate is negative outside
and {\it inside} the oasis ({\it i.e.} $\langle U \rangle < -U_c$),
then the only possible state is extinction at any velocity.
If there is positive growth inside the oasis, but negative
outside, a localized population can survive in the oasis,
but only for small enough wind velocities $v$.
Extended states are present for a small positive growth rate
in the desert ($\langle U \rangle >0$). In this case  
localized and delocalized states coexist for small velocities
(``mixed phase''),
while at large velocities all eigenstates are extended,
as shown in figure {\protect{\ref{delocalization}}}.
The ground state becomes delocalized at the critical
velocity $v_c$ which marks the phase boundary
between the mixed and the extended phase.
\label{phasediagram}}
\end{figure}

A special (universal) behavior is expected for the spatially 
average growth rate $\langle U \rangle  = 0$.
In this case the delocalization and extinction velocities coincide.
Figure \ref{phasediagram} summarizes the different scenarios
in a sketch of the phase diagram for large systems with fixed well depth 
$U_0 \equiv (a+\epsilon a)$, tuning the drift velocity, and 
the average growth rate $ \langle U \rangle = -\epsilon a $.
Also shown in figure \ref{phasediagram} is a horizontal 
transition line at $\langle U \rangle = 0 $ separating a small 
velocity
region ($\epsilon < 0$) where localized modes dominate the steady
state bacterial population, from one ($\epsilon >0$)
containing a mixture of localized and extended states.
The experimental signature of this interesting transition,
(which could be accessed by increasing the light intensity
for photosynthetic bacteria
at fixed convection velocity) will be discussed in a future publication
\cite{long-paper}.
It is of course also possible to drive a population {\it extinct}
at zero velocity simply by lowering the average growth rate. This 
special transition at $\langle U \rangle = -U_c$ is indicated at the 
bottom of figure \ref{phasediagram}.
In section \ref{linear} we give details of the analysis of the 
one dimensional linearized problem for infinite and finite systems
with periodic boundary conditions.
In section \ref{nonlinear} some effects
of the nonlinear term are discussed, 
especially for experiments near the extinction transition,
and in section \ref{2dim} the two dimensional case is discussed.
The appendices contain some details on the analytic computation 
of finite size effects (Appendix \ref{app:finite-syst}), 
a brief discussion of a lattice model \cite{nadav}
corresponding to the analytic continuum theory 
(Apppendix \ref{app:lattice}),
and a discussion of dimensionless 
quantities measurable in experiments 
(Appendix \ref{app:dimensionless-quantities}).

\section{Linearized Growth in one Dimension}
\label{linear}

If the left and right eigenfunctions $\phi^{R,L}_n({\mathbf x})$ 
are localized
({\it i.e.} if the convection velocity is small enough,
so that $\phi^{R,L}_n({\mathbf x})$ decays exponentially
with the distance from the oasis),
one may eliminate the convective term in \Eq{Liouville}
via the transformation
\be
\label{gauge-trf}
\phi^{R,L}_n({\mathbf x}) = \exp(\pm {\mathbf v} 
\cdot {\mathbf x}/2 D) \psi_n({\mathbf x})
\ee
($+$ refers to the right eigenvectors and $-$ to the left 
eigenvectors). The eigenvalue equation associated with 
the linearized growth operator (\ref{linearized-equation}) 
becomes Hermitian \cite{decay}
\be
\label{linear-equation-of-motion}
\Gamma_n \psi_n({\mathbf x}) =  
D \nabla^2 \psi_n({\mathbf x}) + U({\mathbf x}) \psi_n({\mathbf x})
-(v^2/4D) \psi_n({\mathbf x}) \, ,
\ee
and is equivalent to the familiar square well potential problem much studied 
in quantum mechanics.
With the identifications 
$a+\epsilon a \equiv U_0/{\hbar}$, 
where $U_0$ is a quantum well depth, and ${\hbar}$ is 
Planck's constant,
$\Gamma_n + \epsilon a +  v^2/4D \equiv | E |/\hbar  $,
where $E$ is a quantum energy level,
and $D \equiv \hbar/2m$, where $m$ is a mass in the equivalent
quantum problem, we can use well known 
quantum mechanical results\cite{Landau,Fluegge}.
The left and right eigenfunctions at finite velocity are then related 
to the eigenstates of the Hermitian problem 
(\ref{linear-equation-of-motion}) via 
the transformation (\ref{gauge-trf}), 
while the eigenvalues undergo a rigid shift
\be 
\label{shift}
\Gamma_n (v) = \Gamma_n (v = 0) - {v^2 \over 4D}.
\ee

\subsection{An Oasis in an Infinite Desert: Localized 
Populations and the Extinction Transition}
\label{linear-localized-1dim}
In an infinite one dimensional system,
localized solutions for $\phi^{R,L}_n(x)$ 
are given by \Eq{gauge-trf} with \cite{timescale}
%
%
\be
\label{1dim-groundstate}
\psi_n(x) = 
\cases{
A_{1,n} \exp(\kappa_n x)\, ,& for $ x<-{W\over2}\, ,$\cr
\ & \cr
B_{1,n} \exp(i k_n x)+ B_{2,n} \exp(-i k_n x) \, ,& for $-{W\over2} < x < {W\over2}\, ,$\cr
\ & \cr
A_{2,n}  \exp( - \kappa_n x) \, ,& for $x > {W\over2} \, ,$\cr
}
\ee
where $A_{1,n},A_{2,n},B_{1,n},$ and $B_{2,n}$ are constant coefficients,
and
\be
\label{kappan}
\kappa_n = \sqrt{ (\Gamma_n(v =0) + \epsilon a)/D } \, ,
\ee
and 
\be
\label{kn}
k_n = \sqrt{ (a-\Gamma_n(v = 0))/D  }\, ,
\ee
as can be seen by substituting the above Ansatz for $\psi(x)$
into \Eq{linear-equation-of-motion}. \Eq{kappan} implies
\be
\Gamma_n(v = 0)=D\kappa_n^2 -\epsilon a \, .
\ee

To compute $\Gamma_n(v = 0)$ one matches
both $\psi_n(x)$ and $\partial \psi_n(x)/\partial x$ at 
$x= \pm W/2$, which determines the coefficients $A_{i,n}, B_{i,n}$, $i=1,2$,
up to an overall multiplicative factor, 
as well as the eigenvalues $\{\Gamma_n (v=0)\}$.  
When solving the Hermitian 
problem one may use the fact that all the eigenfunctions 
admit a well defined parity, i.e., they are odd or even under the 
transformation $x \rightarrow -x$. Even integers $n$ correspond to 
bound eigenstates with even parity, where  $\psi_n(x)$, 
is symmetric under $x \rightarrow -x$. In such a case 
one obtains the eigenvalue equation for the quantity 
\be
\zeta_n(\Gamma_n) \equiv k_n W/2 = W/2  \sqrt{(a-\Gamma_n (v=0))/D} \,,
\ee
namely,
\be
\label{even-eigenvalue}
\cot(\zeta_n) = \zeta_n {\bar x}/\sqrt{1 - (\zeta_n {\bar x})^2 } \, ,
\ee
(which is equivalent to $\cot(k_n W/2) = k_n/\kappa_n$),
with 
\be
\bar{x}=2\sqrt{D/(a+\epsilon a)}/W=2/(\sqrt{k_n^2 + \kappa_n^2} W) \, .
\ee
The dimensionless parameter ${\bar x}$ measures the ratio of 
kinetic to potential energy in the equivalent quantum problem.
For bound eigenstates with odd parity $\psi(x)$,
($\psi(x)$ antisymmetric under $x \rightarrow -x$, denoted by odd $n$ )
one obtains
\be
\label{odd-eigenvalue}
\cot(\zeta_n) = -\sqrt{1 - (\zeta_n {\bar x})^2 }/(\zeta_n {\bar x}) \, .
\ee
The highest eigenvalue (with its
nodeless, positive eigenfunction $\psi_0(x)$),
corresponds to the largest growth rate $Re\{ \Gamma_0 \}$, and is 
therefore expected to dominate the system in most cases at long times,
as seen from \Eq{time-evolution}.
Its eigenvalue condition (\ref{even-eigenvalue}) for $n=0$ leads 
to \cite{Landau} 
\be
\Gamma_0(v=0) + \epsilon a  = (a + \epsilon a) f(\bar{x}) \, ,
\label{Gamma_0} 
\ee
where $f({\bar{x}})$ is a monotonically decreasing function such that
$f(\bar{x}) \simeq 1 - \pi^2 \bar{x}^2/4$ for $\bar{x} \ll 1$, and 
$f(\bar{x}) \simeq 1/\bar{x}^2$
for $\bar{x} \gg 1$.
Upon inserting \Eq{Gamma_0} into the expressions for $\kappa_n$
and $k_n$ one obtains
\be
\kappa_0 = \sqrt{ (a + \epsilon a) f(\bar{x})/D }
\ee
and 
\be
k_0 = \sqrt{ (( a+ \epsilon a) (1- f(\bar{x})))/D }\, .
\ee
When $\bar{x} \ll 1$, the ``potential well'' is very deep, and one finds
the usual particle in a box result for $\Gamma_0(v)$, 
with correction term proportional
to $v^2$ arising from the change of variables (\ref{gauge-trf}) 
and (\ref{shift}), namely
\be
\Gamma_0(v) \simeq (v_c^2 - v^2)/(4 D) =
a - D \pi^2/W^2 - v^2/4D \, ,
\ee
with 
\be
v_c=2 D \sqrt{(a/D- \pi^2/W^2)} \, ,
\ee
\be
\kappa_0 = \sqrt{(a+ \epsilon a)/D-\pi^2/W^2}\, 
\ee
and
\be
k_0=\pi/W \, .
\ee
For $\bar{x} \gg 1$ one finds 
\be
\Gamma_0(v) \simeq (v_c^2 - v^2)/(4 D) =
(a + \epsilon a)^2 (W/2)^2/D - 
\epsilon a - D (v/2D)^2 
\ee
with 
\be
v_c = 2D \sqrt{((a + \epsilon a) W/2)^2/D^2 - \epsilon a/D} \, ,
\ee
\be
\kappa_0=\sqrt{(a + \epsilon a)^2 (W/2)^2/D^2}
\ee
and 
\be
k_0= \sqrt{ (a+\epsilon a) (1- 1/\bar{x}^2)/D} \, .
\ee

If we take as an effective diffusion constant for motile bacteria 
$D=6\cdot 10^{-6} {\text cm}^2/{\text sec}$, and a growth rate 
$a = 10^{-3}$ /sec in the oasis and a much smaller
growth rate outside ($0<|\epsilon| << 1$),
we get for a $W=2 {\text cm}$ diameter oasis,
$\bar{x} = 0.077 \ll 1$, and an ``extinction velocity''
$v_{c} \simeq 1.5 \mu/{\text sec}$, which is comparable 
to the Fisher wave velocity \cite{murray} in the oasis,
$v_F =2 \sqrt{a D} = 1.5\mu/{\text sec}$.

\subsection{Finite Size Effects and the Delocalization Transition}
An experimental finite system with periodic boundary conditions
is depicted in figure \ref{setup}.
The eigenvalue equation for $\kappa_n(\Gamma_n)$ and $k_n(\Gamma_n)$
of the corresponding linearized problem
of an oasis of width $W$ in a finite desert
of extent $L > W$ (with $L$ being the
mean circumference of the circular
region in figure \ref{setup}),
is obtained using the Ansatz in \Eq{gauge-trf} with
\be
\label{1-dim-eigenstate-finite-syst}
\psi_n(x) = 
\cases{
(A_{1,n} \exp(\kappa_n x) + A_{2,n} \exp(-\kappa_n x)) \, ,& 
for $ -{L\over 2} < x < -{W\over 2}\, ,$\cr \ & \cr
(B_{1,n}\exp(i k_n x)+B_{2,n}\exp( -i k_n x)) \, ,& 
for $-{W\over2} < x < {W\over2}\, ,$\cr
\ & \cr
(C_{1,n} \exp(\kappa_n x) + C_{2,n} \exp(-\kappa_n x)) \, ,& 
for ${W\over2} < x < {L\over 2} \, ,$\cr
}
\ee
and matching $\phi^{R,L}_n(x)$ and $d \phi^{R,L}_n(x)/dx$ at 
the edges of the well and at the edges of the sample (imposing 
periodic boundary conditions).
One finds the eigenvalue equation\cite{Hatano} 
(see also \cite{contact-Hatano})
\bea
\label{finite-syst}
\nonumber
&2& k_n \kappa_n ( cosh( L v/(2D)) - cos(k_n W) cosh(\kappa_n(L-W))) \\
&+& (k^2_n- \kappa^2_n) sin(k_n W) sinh(\kappa_n( L-W)) =0 \, .
\eea
For $v/(2D) < Re\{\kappa_n \}$, and large $L$,
equation (\ref{finite-syst}) yields
\bea
& & \exp(L (v/(2D)-\kappa_n)) = \\
\nonumber
& & \cos( k_n  W) - k_n/(2 \kappa_n) \cdot 
(1 - (\kappa_n/k_n)^2) \sin(k_n W) \, .
\label{fin-syst-localized}
\eea
The left hand side vanishes in the limit $ L \rightarrow \infty $,
and the equation reduces to the bound state equations of a single
square well (\Eq{even-eigenvalue} for even parity
solutions and \Eq{odd-eigenvalue} 
for odd solutions).
For finite $L$, and $v/2D < Re \{ \kappa_n \}$,
the deviation of the ``localized'' solutions $\kappa_n$ and $k_n$ from 
their $v=0$ values for small $v$ is exponentially small in L \cite{Hatano}.
These ``localized''  or ``bound state'' solutions, are
characterized by an exponential 
decay of the bacterial density in the desert with a correlation
length 
\be
\xi_n \sim (Re\{\kappa_n\}-v/(2D))^{-1}
\ee
and 
$\Gamma_n(v)$ strictly real.
However, ``delocalized'' or ``scattering'' solutions also exist, with
$\Gamma_n(v)$ complex,
and nontrivial dependence of $\kappa_n$ and $k_n$ on $v$, 
even in the limit of large $L$.
As the velocity is increased, the $n$th localized eigenstate 
becomes delocalized ($\xi_n \rightarrow \infty$) at the critical 
delocalization velocity $v_n^*$ given in an infinite system by
\be
v_n^* = 2 D Re\{\kappa_n \} \, .
\ee
This implies
\be
\label{nu}
\xi_n \sim 1/(v_n^* -v )^\nu
\ee
with the (universal) critical exponent $\nu=1$.
We saw that with increasing velocity, the growth rate
$Re \{ \Gamma_n(v) \} $ for a given eigenstate decreases. 
It becomes negative above the corresponding extinction velocity
$v_{nc}$.
We therefore expect that the delocalization transition for the 
``ground state'' (which tends to dominate the long time 
behavior) can be observed only if $v_0^* \le v_{0c}$.
In the following we discuss the three desert scenarios,
$\epsilon >0 $, $\epsilon < 0$, and $\epsilon =0$.

(1) For $\epsilon >0$ (a ``deadly'' desert), of big enough size $L$, 
one finds that all delocalized states 
die out exponentially with time ($v_{nc} < v_n^*$).
The population is localized around the oasis at small $v$
and extinct at high $v$.
Figure \ref{Gamma-v-series} 
is a plot of the eigenvalues $\Gamma(v)$ 
in the complex plane for this case, as derived 
for the lattice model discussed in Appendix \ref{app:lattice}.
The lower part of figure \ref{extinction} shows 
a series of profiles of the ground state
eigenfunction close to the extinction transition.
In small enough systems,
such that the total effective growth
rate $\int_0^L U(x) dx$ is positive, delocalized states 
can actually have a positive growth rate even for $\epsilon >0$.
(See also \Eq{Gamma-systemsize} in Appendix \ref{app:finite-syst}, 
with $\delta\kappa \sim O(1/L)$,
and ${\bar \kappa} \sim O(1/L)$ .)
In this case, the system is small enough so that the bacteria
can traverse the desert quickly, and on average won't die before 
reenterring the oasis in a circular pipe. 

\begin{figure}
\epsfxsize=5truein \vbox{\vskip 0.15truein\hskip 1truein
\epsffile{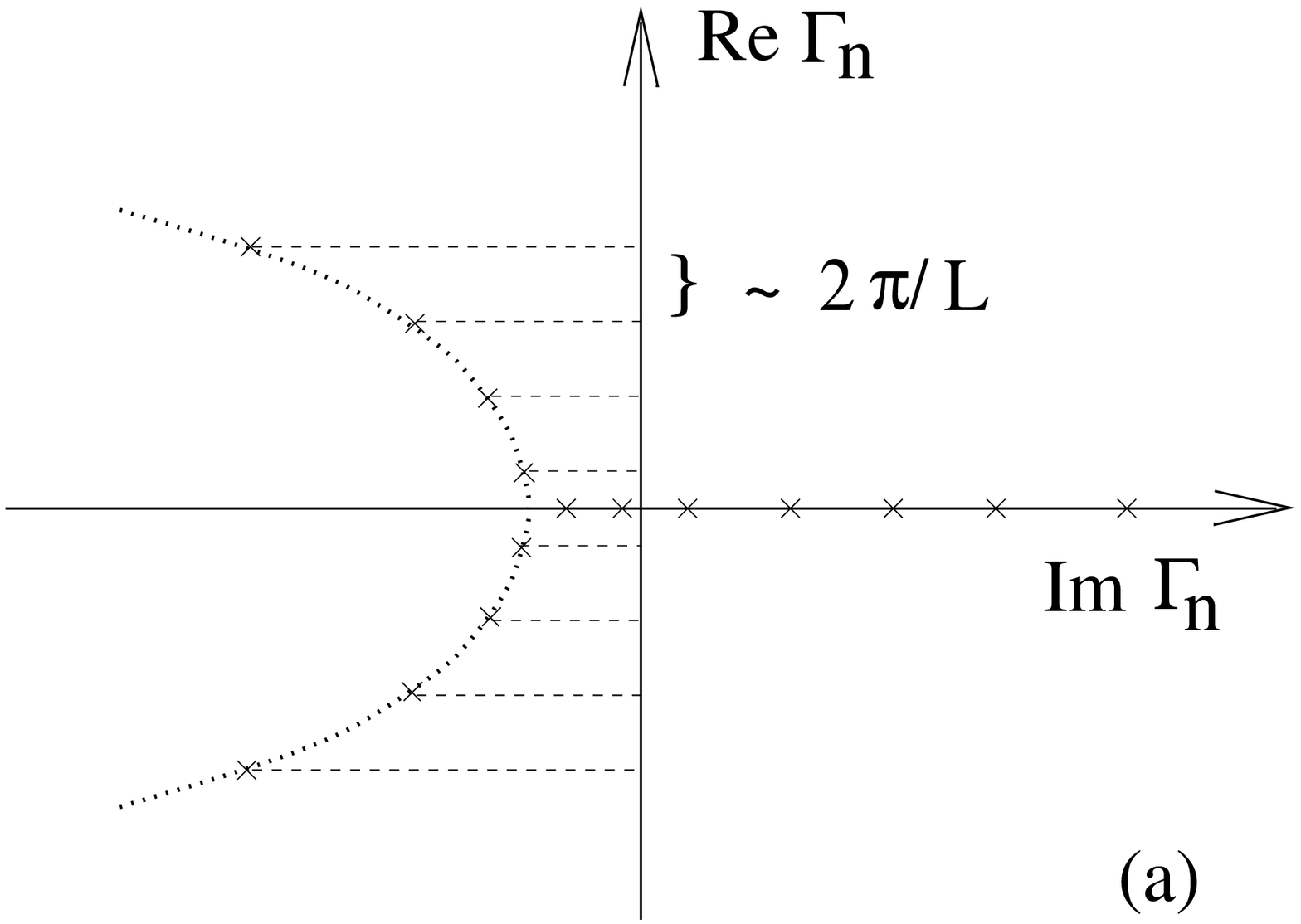}}
\epsfxsize=5truein \vbox{\vskip 0.15truein\hskip 1truein
\epsffile{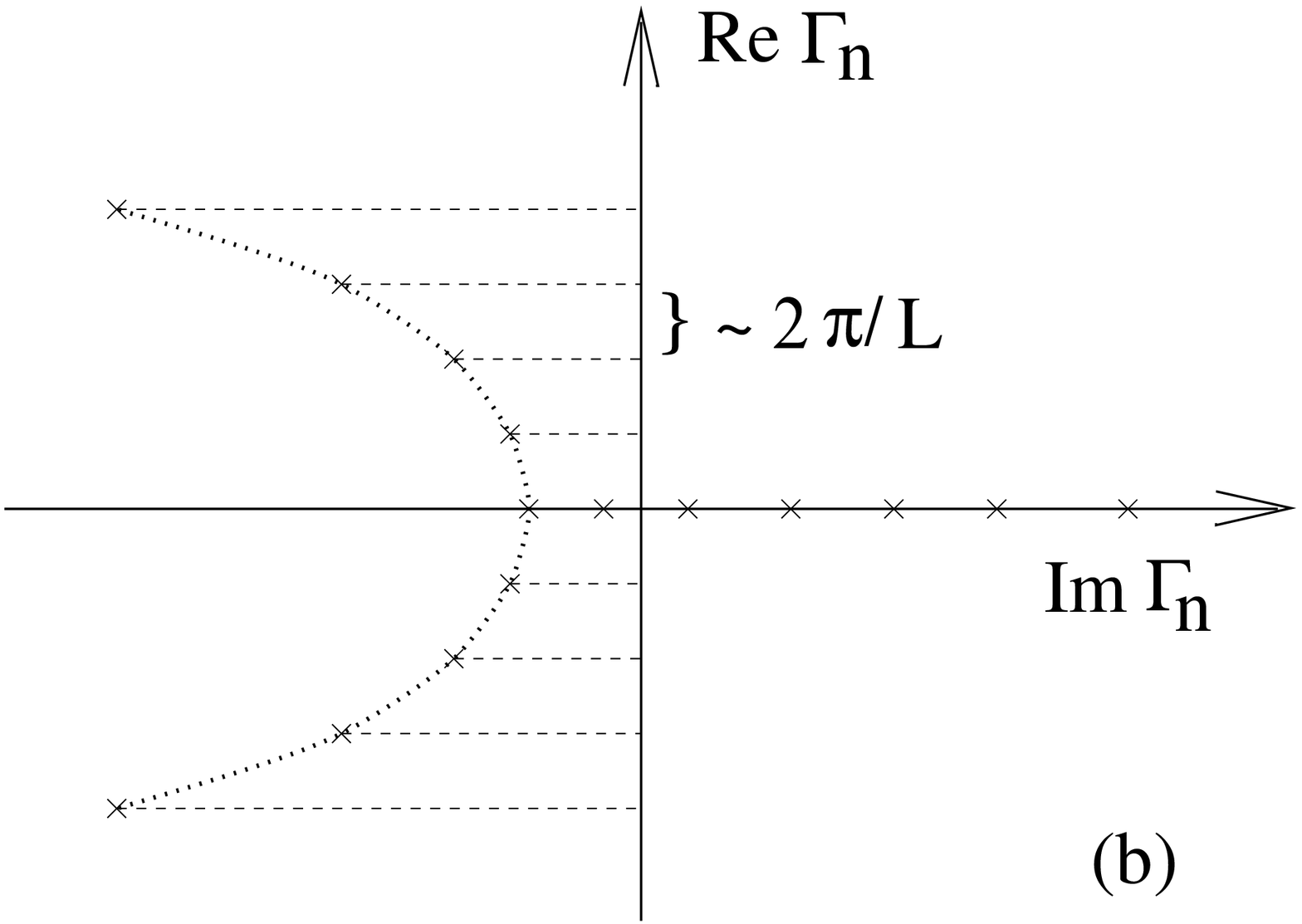}}
\medskip
\caption{Sketches of complex nonhermitian eigenvalue spectra
at two different velocities:
(a) velocity $v_a$ with $v_7^* < v_a <v_6^*$, 
where $n=6$ denotes an even eigenstate, {\it i.e.} all eigenstates 
with $n>6$ are delocalized and those with $n\le 6 $ are localized;
and (b) at a higher velocity $v_b > v_a$ with 
$v_6^* < v_b < v_5^*$ 
where $n=5$ denotes an odd eigenstate. Upon increasing the velocity
from $v_a$ to $v_b$ the $n=6$ eigenstate becomes delocalized at
$v=v^*_6$. 
\label{Gamma-figure}}
\end{figure}

(2) If $\epsilon < 0$, delocalized states should be 
observable even for very large systems, because the ``desert'' can 
support modest growth, although at a much smaller 
rate than in the oasis if $|\epsilon| \ll 1$.
Growing delocalized eigenstates are present, 
even for $v=0$, and the population is a superposition of fast growing 
localized states and more slowly growing delocalized ones.
As a drift velocity $v$ increases, the n'th localized eigenstate
delocalizes at $v=v^*_n$ with a positive growth rate 
$Re{\Gamma_n(v^*_n)}$ 
(i.e. $v^*_n < v_{nc}$). 
This case allows for an experimental observation of the delocalization
transition: as the velocity is increased, more and more eigenstates
delocalize. 
The eigenvalue spectrum for two different values of $v$ 
is shown in figure \ref{Gamma-figure} (a) and (b).
We can see that the spectrum at the delocalization
transition is slightly different depending on whether an ``even''
eigenfunction or an ``odd'' eigenfunction is about to
delocalize next (``even'' and ``odd'' are to be understood
in the sense explained in section \ref{linear-localized-1dim}).
If an odd eigenfunction is about to delocalize next, there exists 
a delocalized state which has a purely real growth rate 
(at the tip of the parabola in the spectrum of figure 
\ref{Gamma-figure}(b)),
while no such state exists when an even eigenfunction is about
to delocalize, as in figure \ref{Gamma-figure}(a).
The essential characteristics of the spectrum
are derived in Appendix \ref{app:finite-syst}.

For experiments, we focus on the delocalization of the ground state,
since it is the fastest growing eigenstate, which dominates the system
near the oasis. 
The ground state delocalizes at the highest delocalization velocity 
$v = v^*_0$, {\it i.e.} for $v> v^*_0$ all states are delocalized. 
The lower part of figure \ref{delocalization} shows a series of profiles 
of the ground state eigenfunction close to delocalization. One sees that
at the delocalization velocity $v^*_0$, the correlation length $\xi_0$
reaches the system size. In an infinite system it diverges 
as in \Eq{nu}. 
Near the delocalization transition,
we expect certain quantities to become
independent of microscopic details of the system. 
These universal quantities will only depend on general properties, 
such as symmetries, dimensions etc.. 
Simple models that only share
these general properties with the experimental system
will be sufficient for precise predictions for these quantities 
near the transition. An example of such a quantity
is the critical exponent $\nu$ in \Eq{nu}.
Details for the square well system and more general
random systems will be presented in a future paper\cite{long-paper}.

(3) For $\epsilon=0$, the growth rate of the bacteria
exactly balances the death rate in the desert, 
they only diffuse.
In this case the delocalization velocity and the extinction
velocities coincide: $v^*_n = v_{nc}$ in an infinite system.
There is again a diverging correlation length 
leading to universal critical behavior near the delocalization 
transition \cite{long-paper}.
In a finite system the average growth rate
$\int_0^L dx U(x)/L$ is positive. 
In small systems,
one therefore expects to see delocalized states with a positive 
growth rate in experiments for this case as well.

These results for one dimension in the limit $L \rightarrow \infty$
are summarized in figure \ref{phasediagram} 
which shows a schematic phase diagram for fixed well depth 
$U_0 \equiv (a+\epsilon a)$, as a function of the convection 
velocity and the average growth rate
$\langle U \rangle \equiv \int_0^L dx U(x)/L$,
in the limit $L \rightarrow \infty$.

\vskip 0.2truein
\begin{figure}
\epsfxsize=5truein \vbox{\vskip 0.15truein\hskip 1truein
\epsffile{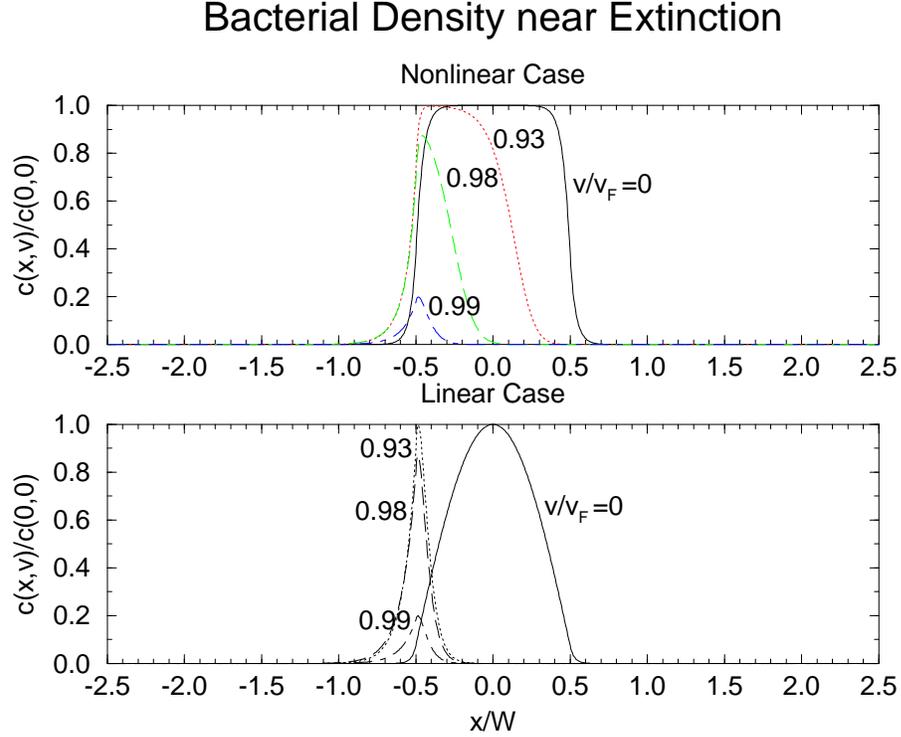}}
\medskip
\caption{A series of steady state population profiles for a system with 
negative average growth rate, at different 
drift velocities, for the linear and the nonlinear case.
The $x$ coordinate is normalized by the width $W$ of the oasis.
In the nonlinear case the populations $c(x,v)$ are divided by the 
population at zero velocity in the middle of the oasis $c(0,0)$.
In the ``linear'' case, we simply assume the validity of 
\Eq{cstar-linear} and rescale the curves such that their maximum
values match those of the nonlinear case.
The profiles were extracted from a lattice model with 1000 sites,
an oasis of width $W=200$ sites, with growth rate $U=0.5$ inside the oasis 
and $U=-0.5$ outside, and diffusion constant $D=30$. 
The velocity parameters 
are taken to be
$v/v_F=0$ (solid line), $v/v_F=0.93$ (dotted), $v/v_F=0.98$ (dashed),
and $v/v_F=0.99$ (dash-dotted).
Close to the extinction transition,
where the population is small, the agreement
between the nonlinear and the linear solution becomes quite good.
The oasis occupies the region $|x/W | < 0.5 $ in the figures.
\label{extinction}}
\end{figure}
\vskip 0.5truein
\begin{figure}
\epsfxsize=5truein \vbox{\vskip 0.15truein\hskip 1truein
\epsffile{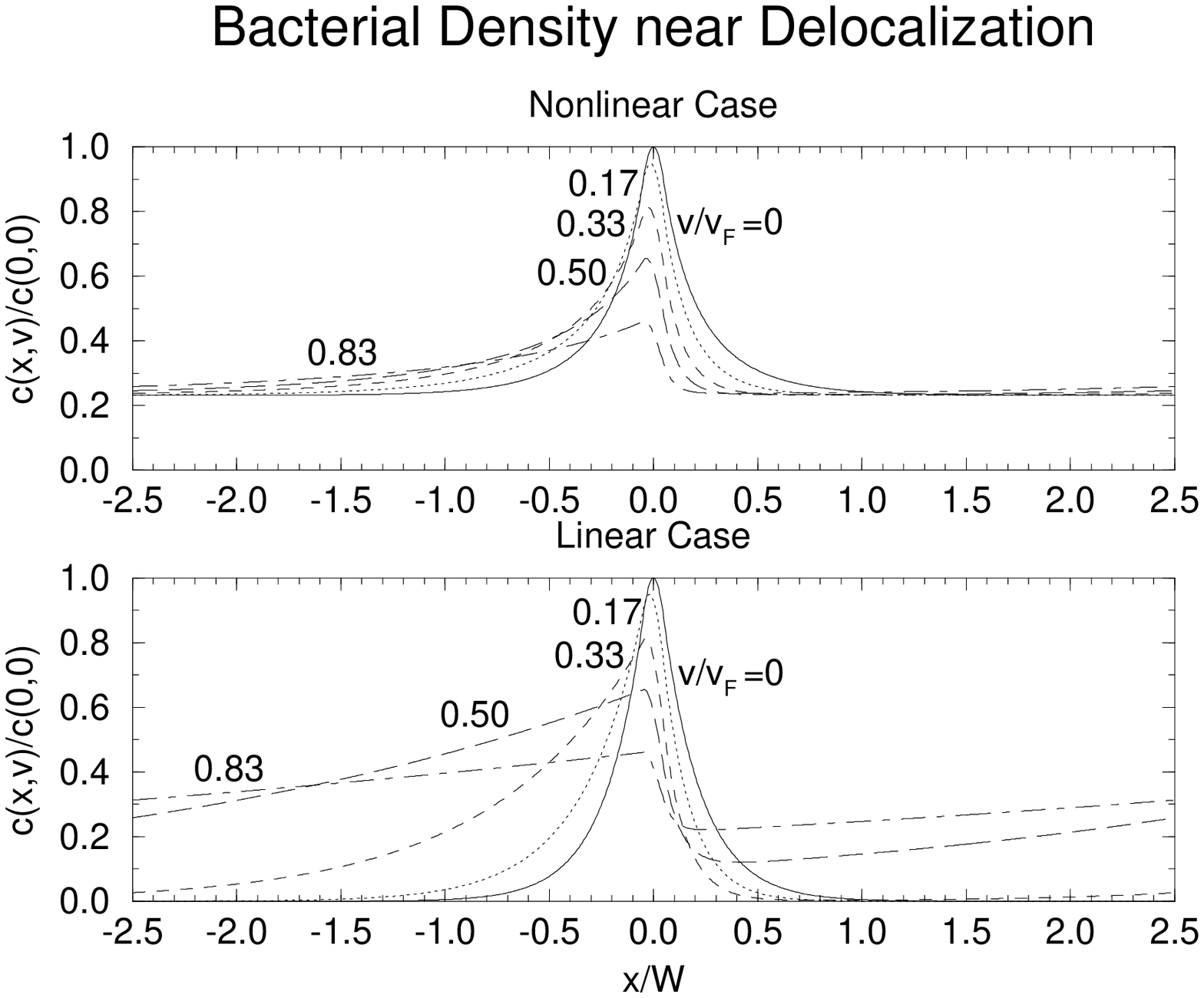}}
\smallskip
\caption{A series of steady state population profiles for a system with 
positive average growth rate, at different 
drift velocities, for the linear and the nonlinear case.
The $x$ coordinate is normalized by the width $W$ of the oasis
$(|x/W| \le 0.5)$.
In the nonlinear case the populations $c(x,v)$ are divided by
$c(0,0)$, the population in the middle of the oasis, at zero velocity.
Curves for the ``linear'' case, were computed as in figure 
{\protect{\ref{extinction}}}.
The profiles were extracted from a lattice model with 1000 sites,
an oasis of width $200$ sites, with growth rate $U=0.1$ inside the oasis 
and $U=1.1$ outside, and large diffusion constant $D=300$. 
The velocity parameters $v/v_F$ are taken to be
$v/v_F=0$ (solid line), $0.17$ (dotted), $0.33$ (dashed),
$0.50$ (long-dashed), and $0.83$ (dash-dotted).
In both cases the population becomes
delocalized at high velocities with increasing velocity 
when the correlation length reaches the size of the system.
\label{delocalization}}
\end{figure}

\subsection{Delta-Function Potential Well}

Results for a delta-function-like oasis in one dimension, can be 
easily derived as a special case of the square well
potential, by taking the limit $\epsilon \rightarrow 0$,
$a \rightarrow \infty$ and $W \rightarrow 0$
with $a W = {\text const} \equiv V_0$.
In this case \Eq{finite-syst} simplifies to 
\be
\kappa (\cosh(Lv/(2D)) - \cosh(\kappa L)) + V_0 \sinh(\kappa L)/D =0 \, ,
\ee
which is the same equation as obtained in \cite{Hatano}.
With the identification $\kappa \equiv (-i) K$ the results derived there
can be applied here: the delocalization picture is the same as in 
figure \ref{Gamma-figure}(a), except that 
there exists only one (even parity) bound state 
solution for $Re\{\kappa\} > v/(2D) $.
Again two critical velocities $v_{c0}$ and $v^*_0$ 
emerge, in accordance to the above discussion for a square well 
potential.

\section{Effects of the Nonlinearity}
\label{nonlinear}
We can also estimate the effects due to the nonlinear term in 
equation~(\ref{equation-of-motion}), which leads to a saturation
of $c(x,t)$ for $ \Gamma_n(v) > 0 $.
The equation of motion becomes 
\be
\partial c(x,t)/\partial t = {\cal L} c(x,t) - b c^2(x,t) \, ,
\ee
with $\cal L$ given by \Eq{Liouville}.
(The coefficient $b$ can be set to 1 by rescaling the density
$c(x,t)$ by $b$, as in Appendix 
\ref{app:dimensionless-quantities}.)
The solution can be expressed in terms of the complete set of 
right eigenstates
for the linear problem with new time dependent coefficients $c_n(t)$ with
\be 
\label{superposition}
c(x,t) = \sum_n c_n(t) \phi^R_n(x)
\ee
and
\be
d c_n(t)/d t = \Gamma_n c_n(t) - \sum_{m,m'} w_{n,mm'} c_m(t) 
c_{m'}(t) \, ,
\ee
where the mode couplings are given by \cite{nadav}
\be
\label{mode-couplings}
w_{n,mm'} = b \int dx \phi_n^L(x) \phi_m^R(x) \phi_{m'}^R(x) \, .
\ee
In general one expects that through the mode couplings the 
fastest growing eigenstate suppresses the growth rate of the other 
eigenstates, provided the corresponding couplings are 
large enough \cite{nadav,long-paper}.
In the mixed phase of figure \ref{phasediagram} we expect that the fastest
growing bound state suppresses the other bound states, and the fastest
growing delocalized state suppresses the other delocalized 
states\cite{long-paper}.

\subsection{Effects of the Nonlinearity at the Extinction Transition}

\begin{figure}
\epsfxsize=5truein \vbox{\vskip 0.15truein\hskip 1truein
\epsffile{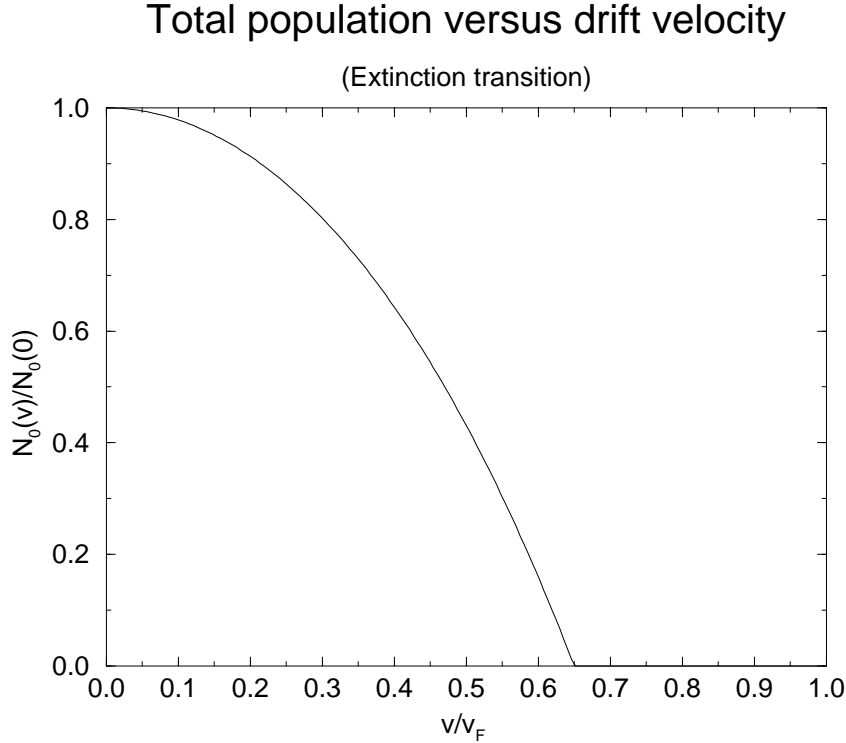}}
\medskip
\caption{Numerical results for the total number of bacteria $N_0(v)$ 
(normalized by the total number at zero velocity $N_0(0)$) as
a function of $v/v_F$, where $v_F= 2 \sqrt{aD}$ is the Fisher wave
velocity. The data were obtained using a 1 dimensional lattice model 
with 100 grid points (see Appendix {\protect{\ref{app:lattice}}}).
The diffusion constant is $D=3.$, the width of the oasis is $W=20$,
the growth rate is -0.9 outside the oasis and 0.1 inside.
The coefficient of the nonlinear term in the equation of motion is
$b \equiv 1$. The average growth rate is $-0.7 < 0$ for this system,
so that $v_{0c} < v^*_0$, and the system goes through the extinction
transition at $v/v_F \simeq 0.65$, above which $N_0(v)$ vanishes.
The approximation $v_c \simeq v_F$ mentioned in the text does not
work quite as well here as in other cases because the
condition ${\bar x} \ll 1$ necessary for this approximation is
violated weakly for the model parameters chosen here.
(${\bar x} \simeq 0.35$).
\label{N0versusv}}
\end{figure}

For $\epsilon >0$ (and large enough systems so that 
$\int_0^L dx U(x) < 0$), with velocity $v$ just below 
the extinction velocity $v_{0c}$,
we expect that 
the growing ground state term $c_0(t) \phi^R_0(x)$ 
dominates the summation (\ref{superposition}) 
for $c(x,t)$.
In a first approximation we neglect
all $c_m(t)$ with $m >0$ and find
\be
d c_0(t)/d t = \Gamma_0 c_0(t) - w_0 c_0^2(t),
\ee
with $w_0 \equiv w_{0,00} = b \int d^dx \phi^L_0(x) (\phi^R_0(x))^2 > 0$.
At long times
\be
c_0(t) = c_0(0) {\frac{\exp(\Gamma_0 t)}{ 1+ c_0(0) (\exp(\Gamma_0 t) -1) 
w_0/\Gamma_0}}
\ee
with asymptotic behavior $\lim_{t \rightarrow \infty} 
c_0(t) = \Gamma_0/w_0 $.
Thus, the steady state population profile should be given approximately 
by 
\be
\label{cstar-linear}
c^*(x) = \Gamma_0 \phi^R_0(x)/w_0 \, ,
\ee
where
\be 
c^*(x) = {\text lim}_{t \rightarrow \infty} c(x,t) \, .
\ee
The total steady state bacterial population $N_0$ is
\be
N_0 = \int dx c^*(x) \, .
\ee
It follows that
\be
N_0 \simeq (\Gamma_0/w_0) \int dx \phi^R_0(x) \, .
\ee
Since $\Gamma_0 \sim (v-v_{c})(v+v_{c})$, and 
$\int dx \phi^R_0(x) \sim {\text const}+ O(v-v_c)$ as 
$v \rightarrow v_c$ from below, 
one finds that near the extinction transition
\be
\label{N0-1dim}
N_0 \sim
\cases{
v_c-v\, ,& for $v \rightarrow v_c^- \, ,$\cr
\ & \cr
0 \, ,& for $v > v_c\, ,$\cr
}
\ee
Figure~\ref{N0versusv} shows the total population $N_0$ as 
a function of the convection velocity $v$
for an oasis in a desert with negative growth rate, which is large
enough so that the average growth rate is negative also.
The data was obtained from a lattice model for the nonlinear 
problem (see Appendix \ref{app:finite-syst}).
The displayed convection velocities
range from  $v = 0$ to velocities larger than the extinction 
velocity $v > v_c$.
One can see that $N_0$ decreases linearly with $v-v_c$ for 
$v \rightarrow v_c^-$, as predicted in \Eq{N0-1dim}.
Figure \ref{extinction} shows a series of population profiles for
increasing velocity, for the linear case
({\it i.e.} taking the ground state of the linear problem
as a solution for the steady state), and for the nonlinear case.
One can see that close to the extinction transition the 
ground state of the linearized growth is an excellent 
approximation, as is to be expected since there the bacterial
density $c(x)$ becomes small, 
so that mode couplings other than $w_0$ 
induced by the nonlinear term 
$b c^2(x)$ become small relative to the linear terms.
\vskip 0.2 truein
\subsection{Limit $v \rightarrow 0$}
In the linearized case,
because \Eq{linear-equation-of-motion} is even in $v$,
we expect $N_0(t) =\int d^dx c({\mathbf x},t) \sim {\text const} 
- {\text const }'\cdot v^2 + O(v^4)$ for small velocities.
In the nonlinear case this symmetry is broken, 
because of the $v$ dependence of the coefficients $w_{nmm'}$
in \Eq{mode-couplings} which arises from the transformation 
(\ref{gauge-trf}).
One then expects in the steady state 
$N_0 \sim {\text const} - 
{\text const }' \cdot |{\bf v}| + O(v^2)$ for small velocities.
The constants depend on $U(x)$ and other nonuniversal details.

\subsection{Effects of the Nonlinearity at the Delocalization
Transition}

Figure \ref{delocalization} shows the same information as figure 
\ref{extinction} for the delocalization transition 
for an oasis in a background with positive growth rate. 
One can see that the linear approximation ({\it i.e.} taking
the ground state of the linear problem as a solution for the
steady state), becomes best for 
high velocities \cite{long-paper}, when the drift 
term in the equation of motion becomes dominant compared
to the nonlinear term.
Furthermore the nonlinear solution at
low velocities is in the ``mixed phase'', {\it i.e.}, it is a 
superposition of extended and localized eigenstates of the system.
A decomposition of the $v=0$ nonlinear solution
into the eigenstates of the system shows \cite{long-paper}
that its leading contributions are from the localized ground 
state eigenfunction and the fastest growing delocalized eigenfunction.
As the drift velocity is increased, the contribution of
localized eigenstates clearly diminishes. Above the delocalization
transition, the steady state is composed only of delocalized modes.

\section{Two dimensions, infinite system size}
\label{2dim}

\begin{figure}
\epsfxsize=4truein \vbox{\vskip 0.15truein\hskip 1.2truein
\epsffile{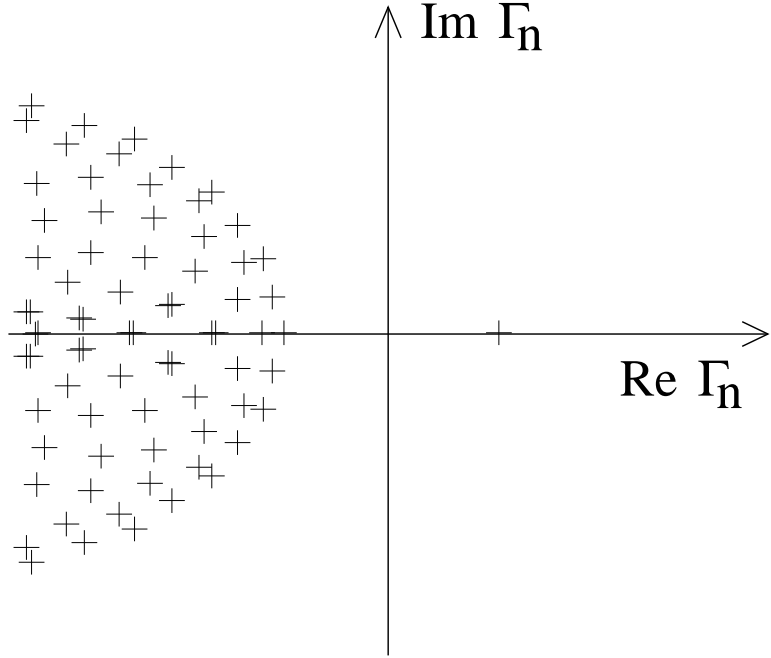}}
\medskip
\caption{Sketch of a complex nonhermitian eigenvalue spectrum in two
dimensions with localized and delocalized eigenstates.
The sketch uses a computation of the spectrum for a delta function 
potential in two dimensions
{\protect\cite{Hatano}}, 
in which case
there is only one bound state (to the right of the
imaginary axis). We expect qualitatively the same result for the
finite two dimensional well discussed in the text, except that
multiple bound states with real eigenvalues will occur, just as
in one dimension.
\label{2-dim-spectrum}}
\end{figure}

Much of the above analysis can be adapted to
two dimensions. We again use the transformation (\ref{gauge-trf})
and consider a circular oasis of diameter $W$
in \Eq{linear-equation-of-motion}.
The analysis is straightforward, so we omit most of the details.
Following reference\cite{Hatano}, the qualitative behavior 
of non-Hermitian eigenvalue spectra for linearized growth
in two dimensions should be as follows:
As in one dimension, localized states associated with 
the potential well will lie in a point spectrum on the 
real axis. Extended states, however, will occupy a dense region 
in the complex plane with a parabolic boundary
(see figure \ref{2-dim-spectrum}). 
With increasing $v$, the localized point spectrum will again
migrate into the continuum. 
In the localized regime
the spatial distribution of bacteria is in the linear approximation 
given by the convection-distorted ground state eigenfunction,
\be
\label{2dim-groundstate}
\phi^{R,L}_0({\mathbf x}) = 
\cases{
C_{1,0} \exp(\pm {\mathbf v \cdot x}/2D)  
        J_0(k_n |{\mathbf x}| )\, ,& for $|{\mathbf x}| < W/2 \, ,$\cr
\ & \cr
C_{2,0}  \exp(\pm{\mathbf v \cdot x}/2D )
 K_0 (\kappa_n |{\mathbf x}| )\, ,& for $|{\mathbf x}| > W/2\, ,$\cr
}
\ee
where $J_0(x)$ and $K_0(x)$ are 
Bessel functions of order zero\cite{Jackson}.
The constants $C_{1,0}$ and $C_{2,0}$ are chosen such that 
$\phi^{R,L}_0({\mathbf x})$ and its radial derivative 
are continuous at $|{\mathbf x}| = W/2$.
The lowest energy eigenvalue is again given by
eq.~(\ref{Gamma_0}), but with different results for $f(\bar{x})$ 
\cite{Landau}:
for $\bar{x} \ll 1$ one finds $f(\bar{x}) \simeq 1- a_1^2 \bar{x}^2$
where $a_1 \simeq 2.405$ is the first zero of $J_0(a)$;
for $\bar{x} \gg 1$ one finds $f(\bar{x}) \simeq \bar{x}^2 \exp(-4 \bar{x}^2)$.
The behavior of the ground state eigenvalue is thus
\be
\Gamma_0 \simeq
a - 4 D (2.405)^2 / W^2 - D v^2/(4 D)
\ee
for $ \bar{x} \ll 1$
and 
\be
\Gamma_0 \simeq D (4/W^2) \exp[-16 D/(W^2(a + \epsilon a ))] - 
D (v/2D)^2
\ee
for $ \bar{x} \gg 1$.
From these results one can compute $v_{0c}$ for the 2 dimensional case.
As in one dimension, one again finds
for $\epsilon >0 $ (``deadly'' oasis),
and large enough systems, that $v_{c0} < v_0^*$,
where $v_0^*/2D \simeq Re\{\kappa_0\}$,
and $v_{0c}$ is the velocity above which the 
ground state growth rate
$\Gamma_0$ becomes negative.

\begin{figure}
\epsfxsize=4truein \vbox{\hskip 1.3truein
\epsffile{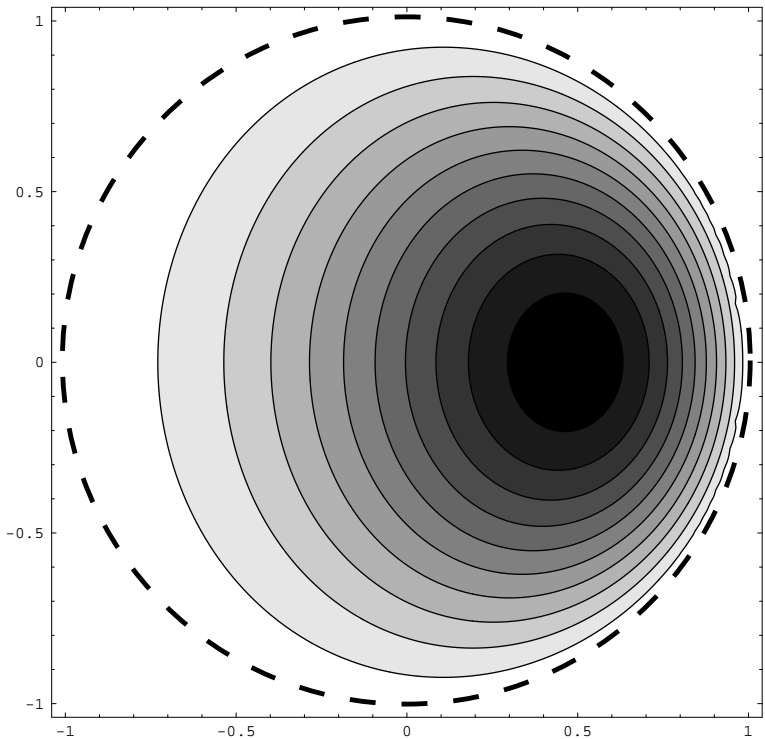}}
\medskip
\caption{Contour plot of the distribution of bacteria in two dimensions
as obtained for the linearized 
problem for convection velocity $v=0.2 \mu/sec < v_c=1.5\mu/sec$ 
(directed to the right),
with $D=6\cdot 10^{-6} {\text cm}^2/{\text sec}$,
and growth rate $a = 10^{-3}$/sec in the oasis (of diameter 
$W=2 {\text cm}$) and a much smaller
growth rate outside ($|\epsilon| = 0.0001 \ll 1$).
The dashed line indicates the circumference of the oasis,
the coordinates are given in units of the radius of the oasis.
At high velocities the linear approximation is expected to give
the steady state generated by 
the full nonlinear equation, provided that
the ground state dominates the long time behavior
in \Eq{superposition}
(see text).
\label{2dim-distribution}}
\end{figure}

Figure~\ref{2dim-distribution} shows a contour plot
for the spatial distribution of bacteria
in the linearized case 
at convection velocity $v$ below $v_{0c}$ 
in 2 dimensions, obtained from
equation~(\ref{2dim-groundstate}).
Although the full nonlinear two dimensional 
problem seems tractable numerically, we expect that
close to the extinction transition the curves for the linearized 
case will approximate the shape of the steady state, because
the ground state dominates sums like that 
in equation (\ref{superposition}).
\appendix

\section{Complex nonhermitian eigenvalue spectra 
for finite system size}
\label{app:finite-syst}
To compute the growth rate $\Gamma_n(v)$ near the
delocalization of the $n$th eigenstate,
we set $ \kappa = v/(2D) + \delta\kappa - i {\bar \kappa} $
where $\lim_{L \rightarrow \infty} \delta\kappa = 0$ \cite{Hatano},
and take $v$ to be close to the corresponding
$v_n^*$. Upon expanding the right hand side of
equation (\ref{fin-syst-localized}) 
in powers of $v-v_n^*$ and $\delta \kappa - i {\bar \kappa}$,
one obtains:
\bea
\nonumber
\exp(-L(\delta \kappa - i {\bar \kappa})) &=& c_n (v-v_n^*)/v + 
O(\delta \kappa - i {\bar \kappa}) \\
&+& O((v-v_n^*)^2)  \, ,
\eea
where $c_n$ is the derivative of the right hand side with respect 
to $v$ at $v=v_n^*$ and $\delta \kappa - i {\bar \kappa} =0$.
For even $n$ one finds $c_n < 0 $, and for odd $n$ one finds $c_n > 0$.
Hence, for even $n$, and $v < v_n^*$, one has to leading order
$ \delta \kappa \simeq \ln [v/(c_n (v-v_n^*))]/L$ and
$ {\bar \kappa} = \pi (2 m + 1)/L$, with $m$ any integer.
For $v > v_n^*$,
$ \delta \kappa \simeq \ln [v/(c_n (v_n^*-v))]/L$ and
$ {\bar \kappa} = 2 \pi m/L$.
Similarly, for odd $n$ and $v < v_n^*$,
$ \delta \kappa \simeq \ln [v/(c_n (v_n^*-v))]/L$ and
$ {\bar \kappa} = 2 \pi m/L$,
and, for $v > v_n^*$, one has
$ \delta \kappa \simeq \ln [v/(c_n (v-v_n^*))]/L$ and
$ {\bar \kappa} = \pi (2 m + 1)/L$.
At $v=v_n^*$ in either case, 
$\delta \kappa \sim O( (\ln L)/L )$ and
${\bar \kappa} \sim O( (\ln L)/L )$ \cite{Hatano}.
For given $v$ and $\kappa$ the value of $\Gamma$ results from 
above definition
of $\kappa$, $\Gamma(v)=D\kappa^2 -\epsilon a - D (v/(2D))^2 $, 
and thus
\bea
\label{Gamma-systemsize}
\nonumber
\Gamma(v) &=& D (\delta\kappa + v/2D)^2 - D {\bar \kappa }^2 -D (v/(2D))^2 
- \epsilon a \\
& & - 2 i D (\delta \kappa+ v/2D)  {\bar \kappa}\, .
\eea
These results are illustrated in figure \ref{Gamma-figure}.

\section{Numerical analysis of a discrete lattice model}
\label{app:lattice}

A discrete lattice model, originally inspired 
by the physics of vortex lines \cite{Hatano},
has proven very helpful for the numerical analysis of our problem.
The corresponding lattice discretization (with lattice constant $l_0$)
of the nonlinear equation
in $d$ dimensions reads \cite{nadav}
\begin{eqnarray}
{dc_{\mathbf x}(t) \over dt}& =& {w \over 2}
\sum_{{\mathbf \nu} = 1}^d 
[e^{{\mathbf g} \cdot {\mathbf e}_{\mathbf \nu}}c_{{\mathbf x} +
{\mathbf e}_{\mathbf \nu}}(t) +
e^{-{\mathbf g}\cdot{\mathbf e}_{\mathbf \nu}} c_{{\mathbf x}-
{\mathbf e}_{\mathbf \nu}}(t) - 2 cosh({\mathbf g} 
\cdot{\mathbf e}_{\mathbf \nu}) c_{\mathbf x}(t) ]
\nonumber \\
&&+  
U({\mathbf x})c_{\mathbf x}(t)-
b c_{{\mathbf x}}^2(t)\;,
\eqnum{1.18}
\label{eq:eighteen}
\end{eqnarray}
where $c_{\mathbf x}(t)$ is the species 
population at the sites $\{{\mathbf x}\}$ of a hypercubic 
lattice, and the  $\{{\mathbf e}_{\mathbf \nu}\}$ are unit lattice vectors. 
Furthermore, $(w \simeq D/\ell_0^2)$, where $D$ is the diffusion constant
of the corresponding continuum model, 
and $g \simeq v \ell_0/(2 D) $, where $v$ is the 
convective flow rate of the continuum model.
$U({\mathbf x})$ and $b$ have the same interpretation 
as in the continuum model \cite{nadav}.
The subtraction in the first term insures that 
$c_{\mathbf x}(t)$ is conserved 
($ {d\over dt}\sum_{\mathbf x}c_{\mathbf x}(t) =0 $)
if $U(x) = b = 0 $.
There are two constraints
on the mesh size or lattice spacing $\ell_0$,
which ensure that the model correctly describes the continuum limit
for a given eigenstate $\phi^R_n$, derived from the condition
that $\ell_0 |\nabla \phi^R_n({\mathbf x})|/\phi^R_n({\mathbf x}) \ll 1$.
For small $v$ one requires
\be
k_n \ell_0 \ll 1 \,\,\,\, {\text{and}} \,\,\,\, \kappa_n \ell_0 \ll 1 \, .
\ee
For high velocities the condition becomes
\be
v \ell_0/(2D) \ll 1 \, ,
\ee
as follows from \Eq{gauge-trf}. The lattice simulations
shown in this paper are well within these limits.

\section{Dimensionless Quantities}
\label{app:dimensionless-quantities}

The equation of motion can be rewritten in dimensionless
form, by introducing rescaled coordinates,
${\mathbf y}\equiv {\mathbf x}/W$, where $W$ is the width $W$ of the 
oasis, and rescaled population densities $\bar c \equiv c/c_s$,
where $c_s$ is roughly the saturation value of the bacterial density
in the oasis (up to $O(\epsilon)$): $c_s = a/b$ (see 
\Eq{equation-of-motion}).
One then obtains 
\bea
\label{dimensionless-equation-of-motion}
\nonumber
(W^2/D) \partial {\bar c}({\mathbf x},t)/\partial t &=&
 \nabla_y^2 {\bar c}({\mathbf x}, t) - 
{\bar {\mathbf v}} \cdot \nabla_y {\bar c}({\mathbf x},t) \\
& & +{\bar v_F}^2 [( U({\mathbf x})/U_0) {\bar c}({\mathbf x}, t)
- {\bar c}^2({\mathbf x}, t)] \, ,
\eea
where ${\bar {\mathbf v}} \equiv {\mathbf v} W/D$ is the dimensionless
rescaled drift velocity,
and ${\bar v_F} \equiv 2 \sqrt{W^2 a/D}$ is the dimensionless
rescaled Fisher wave velocity $v_F = 2\sqrt{D a}$ in the oasis, 
which gives the speed at which a Fisher wave \cite{murray} 
would propagate in the oasis. The velocity $v_F$ also gives a rough 
estimate for
the velocity at which the extinction transition takes place,
in the appropriate parameter regime ($\epsilon < 0$) of the phase 
diagram of figure \ref{phasediagram}.
The basic time scale of the system is set by the
diffusion time $W^2/D$, which is the time it takes a bacterium 
to diffuse across 
the oasis.

\vspace{1cm}
\bigskip
Acknowledgements: It is a pleasure to acknowledge conversations
with Jim Shapiro and Elena Budrene.
This work is supported by the National Science Foundation 
through Grant No. DMR97-14725, and by the Harvard Materials Research
Science and Engineering Center through Grant No. DMR94-00396.
One of us (K.D.) gratefully acknowledges support from the
Society of Fellows of Harvard University.


\begin{references}
\bibitem{ref-frame}
For a moving mask,
as in figure~\ref{setup}, the Fisher equation reads
$\partial c({\mathbf x}',t)/\partial t =
D (\nabla')^2 c({\mathbf x}', t) +
U({\mathbf x}'-{\mathbf v} t) c({\mathbf x}', t)
- b c^2({\mathbf x}', t)$, where ${\mathbf x}'$ is the 
coordinate in the laboratory frame where the growth 
medium is fixed. Upon transforming 
to new spatial coordinates ${\mathbf x} = {\mathbf x'} - {\mathbf v} t$,
in which the mask remains fixed, and then making the 
replacement $c({\mathbf x} + {\mathbf v} t, t) 
\rightarrow c({\mathbf x}, t)$, we obtain 
equation~(\ref{equation-of-motion}).
\bibitem{decay} 
For the transformation~(\protect{\ref{gauge-trf}}) to make physical sense,
the eigenfunctions $\psi_n(\mathbf{x})$ of (\ref{linear-equation-of-motion})
must decay sufficiently
fast at infinity. We check this condition at the end of the
calculation; when it is violated, the eigenvalues become complex.
\end{references}
\end{document}